\documentclass[fp, twocolumn]{jpsj3}

\usepackage{amsmath}
\usepackage{amssymb}
\usepackage{amsfonts}
\usepackage{ascmac}
\usepackage{bm}
\usepackage{calc}
\usepackage{capt-of}
\usepackage[]{circuitikz}
\usepackage{comment}
\usepackage{float}
\usepackage{framed}
\usepackage{graphicx}
\usepackage{mathrsfs}
\usepackage{mdframed}
\usepackage{siunitx}
\usepackage{tabularx}
\usepackage{tikz}
\usepackage{txfonts}
\usepackage{url}
\usepackage{cuted}

\title{
  Thermodynamics of Computation for CMOS NAND Gate
}

\author{
  Daigo Yoshino and Yasuhiro Tokura 
}

\inst{
  Faculty of Pure and Applied Sciences, University of Tsukuba, Tsukuba, Ibaraki 305-8571, Japan
}

\abst{
  Understanding how much energy is needed and dissipated as heat for a given computational system and for a given program is a physically interesting and practically important problem. However, the thermodynamic costs of computational systems are only partially understood. In this paper, we focus on a specific logic gate, the CMOS NAND gate, operating in the sub-threshold region and analyze the dissipated heat from two aspects. One is the general Landauer bound, which is the change in entropy of the computational system, and the other is a cost that depends on the difference between the initial and steady-state distributions of the system. We find that the general Landauer bound is the same order for different inputs to the gate, but that the another cost has partially different order due to the difference between the initial and steady-state distributions over output logical states. We also investigate the interplay between the costs, time scale, and reliability of the process and find that for different inputs, there is not always a trade-off between reliability and dissipation of computations.
}

\begin{document}
\maketitle
\ 
\vspace{-12pt}
\section{Introduction}
All real-world computers need energy to run a given program and most of the energy is dissipated into the external environment as heat. This implies that there is some connection between the thermodynamic properties of nature and the performance of the computer. This is also an important consideration for data centers whose energy consumption increases as the demand for data processing grows \cite{andrae2015global}. In order to solve this physically interesting and practically important problem, one should answer a fundamental question: for a given computational system implemented physically and for a given computational task, how much energy is needed and dissipated as heat? So far, the answer to this question has only been partially given.

In 1961, Landauer \cite{5392446} showed that there is a lower bound for heat dissipated into the thermal environment when the system reliably runs the so-called ``erasure process", a logically irreversible map on 1 bit. Today, this is known as ``Landauer's principle" and the minimum value of dissipation is $k_{\rm B}T \ln 2$, where $k_{\rm B}$ is the Boltzmann constant, and $T$ is the temperature of the thermal reservoir coupled to the system. This minimum value is called the Landauer limit or the Landauer bound. The bound is so tiny that one might think it would have no practical implications. Indeed, it is only about $3.0 \times 10^{-21}\,$J at room temperature. However, this bound can be achieved only if the process is quasistatically performed. More energy may be dissipated in running the erasure process at a practical level, such as in a finite-time erasure process. Thanks to the development of nonequilibrium statistical mechanics, in particular, the birth of the new subfield, called ``stochastic thermodynamics", we have been able to address those problems \cite{wolpert2019stochastic}. Stochastic thermodynamics can describe the system driven out of equilibrium, by using the notions of probability theory and information theory. It has provided us with a fundamental understanding of the thermodynamic irreversibility of physical systems via the fluctuation theorems \cite{jarzynski1997nonequilibrium, crooks1999entropy, collin2005verification, murashita2014nonequilibrium, seifert2012stochastic}, thermodynamic speed limits \cite{shiraishi2018speed, owen2019number}, thermodynamic uncertainty relations \cite{barato2015thermodynamic, gingrich2016dissipation, horowitz2020thermodynamic} and more. Moreover, it has been helpful in understanding the thermodynamic properties of computation, such as the finite-time erasure process \cite{diana2013finite, zulkowski2014optimal, proesmans2020finite, zhen2021universal} and the difference between thermodynamic reversibility and logical reversibility \cite{sagawa2014thermodynamic}.

However, most of the earlier results were on the erasure process \cite{bennett2003notes, berut2012experimental, diana2013finite, zulkowski2014optimal, PhysRevLett.117.200601, proesmans2020finite, zhen2021universal, ma2022minimal}. Although the Landauer bound can be extended to a general computation such as in a logic circuit \cite{wolpert2020thermodynamics}, Brownian computer \cite{bennett1982thermodynamics, strasberg2015thermodynamics, utsumi2022computation}, finite automaton \cite{chu2018thermodynamically}, or Turing machine \cite{strasberg2015thermodynamics, kolchinsky2020thermodynamic}, these results are based on the abstract models and ideal settings. To evaluate the thermodynamic cost for a physical computer, we need to model the physical device and examine discrepancies between the actual cost and the earlier results.

In this paper, we focus on a specific logical computation, the NAND gate, and analyze the dissipated heat from two aspects. One is the general Landauer bound, which is the change in entropy of the computational system, and the other is a cost that depends on the difference between the initial and steady-state distributions of the system. The NAND gate is one of the basic gates of today's complementary metal-oxide semiconductor (CMOS) based computer technologies. We construct a model of the NAND gate by using a thermodynamically consistent physical theory for nonlinear circuits such as diodes and MOS transistors \cite{freitas2021stochastic}. Then, we analyze the general Landauer bound and the above stated cost numerically by using the Gillespie stochastic simulation algorithm (GSSA), which can sample a single trajectory of the microscopic state. Taking the results, we discuss the logical irreversibility of the gate, the initial logical state dependency of the costs and the gap between the costs and the actual dissipated heat. We also investigate the interplay between the costs, time scale and reliability of computations for different inputs to the gate.


\section{Model of CMOS NAND Gate}
\subsection{CMOS NAND Gate}
\vspace{-2pt}
\begin{figure}[tbp]
\begin{center}
    \includegraphics[scale=0.55]{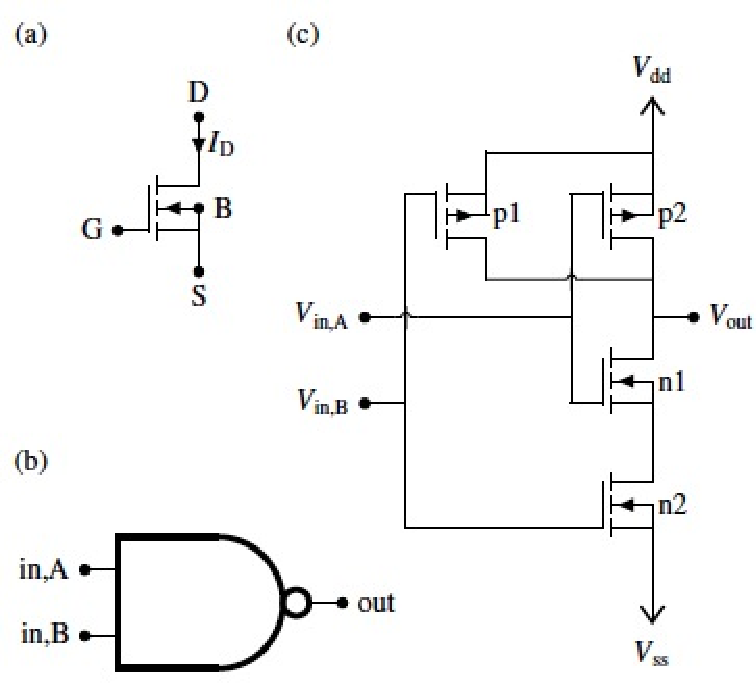}
    \end{center}
  \caption{(a) Symbol for enhancement-type nMOS transistor: B D G and S stand for bulk, drain, ground, and source terminals. (b) Symbol for the NAND gate. (c) Circuit diagram of a typical CMOS NAND gate. p1 and p2 represent enhancement-type pMOS transistors and n1 and n2 represent enhancement-type nMOS ones.}
  \vspace{7pt}
  \label{fig:cmosnand}
\end{figure}
The NAND gate is one of the basic logic gates for implementing a given Boolean function and it is the logical negation of the AND gate. 
Since the NAND gate by itself can be used to implement any Boolean function, it is also called the universal gate.
A typical CMOS NAND gate consists of four MOS transistors, i.e., two n-type MOS (nMOS) transistors and two p-type MOS (pMOS) transistors. These two pairs of transistors are complementarily arranged (see Fig. \ref{fig:cmosnand}). The logical states of the input and output are encoded in the values of the voltages, $V_{\rm in,A}$, $V_{\rm in,B}$, and $V_{\rm out}$.
\vspace{2pt}
\subsection{Basic Setup}
\vspace{-2pt}
Here, we construct a model of the CMOS NAND gate operating in the sub-threshold region based on the framework of the stochastic thermodynamics of nonlinear circuits. \cite{freitas2021stochastic} The sub-threshold, or weak inversion, region is one of the regions in the current-voltage curve of MOS transistors where the gate voltage $V_{\rm g}$ is lower than a certain threshold voltage $V_{\rm th}$, i.e., $V_{\rm g} < V_{\rm th}$ (see Ref. \citen{tsividis2012operation} for details). Computational systems operating in this region are expected to used for low-energy computing. The average drain current has an exponential dependence with respect to the gate source voltage. For nMOS transistors, if the bulk (B) terminal connects to the source (S) terminal in Fig. \ref{fig:cmosnand} (a), the average current is given by
\vspace{-12pt}\\
\begin{align}
  \langle I_{\rm D} \rangle 
  = I_0\,e^{(V_{\rm G} - V_{\rm S} - V_{\rm th})/(nV_{\rm T})}
  \left(1-e^{-(V_{\rm D} - V_{\rm S})/V_{\rm T}}\right),
\end{align}
\vspace{-12pt}\\
where $I_0$ is a specific current, $V_{\rm T}=k_{\rm B}T/q_{\rm e}$ is the thermal voltage with an elementary charge $q_{\rm e}$, and $n\geq 1$ is the slope factor. Furthermore, in the sub-threshold region, the drain current contains shot noise \cite{261888}, which is noise that originates from the discrete properties of the carriers. Shot noise can be modeled as a Poisson process. Ref. \citen{freitas2021stochastic} shows that if the transport of an elementary charge between conductors follows a bi-directional Poisson process (BPP), which means that both the forward and reverse directions of the transport follow a Poisson process, one can determine the transition rates satisfying the local detailed balance (LDB) condition from the I-V curve of the device even if the voltages of conductors change before and after the transition. For fixed applied voltages, the transition rates of transporting an elementary charge through the channel of the nMOS transistor are given by
\vspace{-9.5pt}\\
\begin{align}
  \lambda_+^{\rm n} &= (I_0/q_{\rm e})\,e^{(V_{\rm G} - V_{\rm S} - V_{\rm th})/(nV_{\rm T})} \\
  \lambda_-^{\rm n} &= (I_0/q_{\rm e})\,e^{(V_{\rm G} - V_{\rm S} - V_{\rm th})/(nV_{\rm T})}
  e^{-(V_{\rm D} - V_{\rm S})/V_{\rm T}},
\end{align}
\vspace{-9.5pt}\\
where the subscripts $\pm$ represent the direction of transport of the charge and one can take the forward direction arbitrarily. For the pMOS transistor, the transition rates are
\vspace{-9.5pt}\\
\begin{align}
  \lambda_+^{\rm p} &= (I_0/q_{\rm e})\,e^{(V_{\rm S} - V_{\rm G} - V_{\rm th})/(nV_{\rm T})} \\
  \lambda_-^{\rm p} &= (I_0/q_{\rm e})\,e^{(V_{\rm S} - V_{\rm G} - V_{\rm th})/(nV_{\rm T})}
  e^{-(V_{\rm S} - V_{\rm D})/V_{\rm T}}.
\end{align}
\vspace{-9.5pt}\\
From these rates, we can confirm the LDB condition for this case. For the case where the applied voltages change before and after the transitions, however, a naive derivation of the rates results in a thermodynamic inconsistency, which means that the LDB condition cannot be satisfied. This problem can be solved by taking the average of the voltages before and after the transition and applying it to the transition rates. 

\begin{figure*}[ht]
\begin{center}
    \includegraphics[scale=0.55]{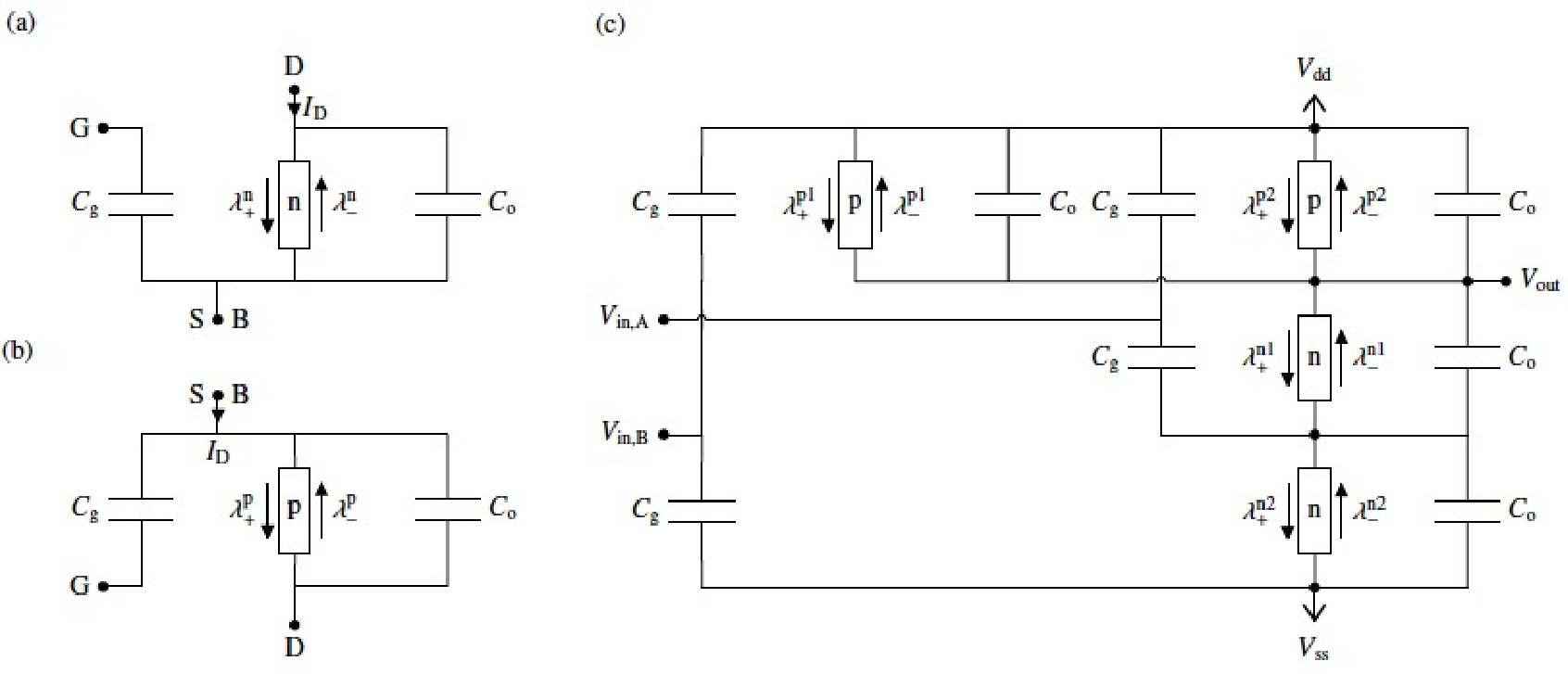}
    \end{center}
  \caption{(a) Model of nMOS transistor. (b) Model of pMOS transistor. (c) Model of CMOS NAND gate}\label{cmosnand_model}
\end{figure*}

As in the case of the CMOS inverter and the full-CMOS probabilistic bit modeled in Ref. \citen{freitas2021stochastic}, we model the MOS transistor as an externally controlled two-terminal device with associated Poisson rates. Figs. \ref{cmosnand_model} (a) and (b) show the circuit diagrams of the MOS transistor model, where the capacitor $C_{\rm g}$ represents the gate-body interface and $C_{\rm o}$ is the output capacitance. 
On the basis of the model for the transistors, we can construct a minimal CMOS NAND gate model operating in the sub-threshold region, as shown in Fig. \ref{cmosnand_model} (c). The model consists of six conductors and four two-terminal devices n1, n2, p1, and p2. Four of the conductors have fixed voltages, namely, $V_{\rm in,A}, V_{\rm in,B}, V_{\rm dd}$ and $V_{\rm ss}$. Hence, the state of the system is determined by the charges of the other conductors, $q_{\rm out}$ and $q_{\rm nn}$. The voltages for these free conductors are respectively written as
\begin{align}
  V_{\rm out}(q_{\rm out},q_{\rm nn})
  &=\dfrac{2+r}{C}q_{\rm out} + \dfrac{1}{C}q_{\rm nn} + V_{\rm cout}\\
  V_{\rm nn}(q_{\rm out},q_{\rm nn})
  &=\dfrac{1}{C}q_{\rm out} + \dfrac{3}{C}q_{\rm nn} + V_{\rm cnn},
\end{align}
where $C = 5C_{\rm o} + 3C_{\rm g}$ and $r=C_{\rm g}/C_{\rm o}$, and
\vspace{-9.5pt}\\
\begin{align}
  V_{\rm cout}&=
  \dfrac{C_{\rm g}}{C}V_{\rm in,A}+\dfrac{2(2C_{\rm o}+C_{\rm g})}{C}V_{\rm dd}+\dfrac{C_{\rm o}}{C}V_{\rm ss}\label{vcout}\\
  V_{\rm cnn}&=
  \dfrac{3C_{\rm g}}{C}V_{\rm in,A}+\dfrac{2C_{\rm o}}{C}V_{\rm dd}+\dfrac{3C_{\rm o}}{C}V_{\rm ss}.\label{vcnn}
\end{align}
\vspace{-9.5pt}\\
The time evolution of the state $\boldsymbol{q}=(q_{\rm out},q_{\rm nn})^{\rm T}$ is stochastically determined from the transition rates of the two-terminal devices. For device n1, the transition rates in the forward and reverse directions are given by
\vspace{-7.5pt}\\
\begin{align}
  \lambda_+^{\rm n1}(q_{\rm out},q_{\rm nn})
  &=t_0^{-1}e^{(V_{\rm in,A}-V_{\rm cnn})/(nV_{\rm T})}
  e^{-(q_{\rm out}+3q_{\rm nn}+q_{\rm e})/(nq_{\rm T})}\\
  \lambda_-^{\rm n1}(q_{\rm out},q_{\rm nn}) 
  &=\lambda_+^{\rm n1}(q_{\rm out}+q_{\rm e},q_{\rm nn}-q_{\rm e}) e^{-(V_{\rm cout}-V_{\rm cnn})/V_{\rm T}}\nonumber\\
  &\quad\times e^{-((1+r)q_{\rm out}-2q_{\rm nn}+((r+3)/2)q_{\rm e})/q_{\rm T}},
\end{align}
\vspace{-7.5pt}\\
where $t_0=(q_{\rm e}/I_0)e^{(V_{\rm th}/(nV_{\rm T}))}$ is the global time scale and $q_{\rm T}=CV_{\rm T}$. As explained above, these quantities are derived by using the average of the voltages before and after the transition. The rates of the other devices n2, p1 and p2 are
\vspace{-7pt}\\
\begin{align}
  \lambda_+^{\rm n2}(q_{\rm out},q_{\rm nn})
  &=t_0^{-1}e^{(V_{\rm in,B}-V_{\rm ss})/(nV_{\rm T})}\\
  \lambda_-^{\rm n2} (q_{\rm out},q_{\rm nn})
  &=\lambda_+^{\rm n2}(q_{\rm out},q_{\rm nn}+q_{\rm e})e^{-(V_{\rm cnn}-V_{\rm ss})/V_{\rm T}} \nonumber\\
  &\quad\times e^{-(q_{\rm out}+3q_{\rm nn}+(3/2)q_{\rm e})/q_{\rm T}}\\
  \lambda_+^{\rm p1}(q_{\rm out},q_{\rm nn})
  &=t_0^{-1}e^{(V_{\rm dd}-V_{\rm in,B})/(nV_{\rm T})}\\
  \lambda_-^{\rm p1}(q_{\rm out},q_{\rm nn})
  &=\lambda_+^{\rm p1}(q_{\rm out}-q_{\rm e},q_{\rm nn})
  e^{-(V_{\rm dd}-V_{\rm cout})/V_{\rm T}} \nonumber\\
  &\quad\times e^{((2+r)q_{\rm out}+q_{\rm nn}-(1+r/2)q_{\rm e})/q_{\rm T}}\\
  \lambda_+^{\rm p2}(q_{\rm out},q_{\rm nn})
  &=t_0^{-1}e^{(V_{\rm dd}-V_{\rm in,A})/(nV_{\rm T})}\\
  \lambda_-^{\rm p2}(q_{\rm out},q_{\rm nn})
  &=\lambda_+^{\rm p2}(q_{\rm out}-q_{\rm e},q_{\rm nn})
  e^{-(V_{\rm dd}-V_{\rm cout})/V_{\rm T}} \nonumber\\
  &\quad\times e^{((2+r)q_{\rm out}+q_{\rm nn}-(1+r/2)q_{\rm e})/q_{\rm T}}.
\end{align}
\vspace{-7pt}\\
Hence, the probability that the state of the system is $\boldsymbol{q}$ at time $t$, which is denoted as $P(q_{\rm out},q_{\rm nn};t)$, follows the following master equation:
\vspace{-8pt}\\
\begin{align}
  &\quad d_tP(q_{\rm out},q_{\rm nn};t)\nonumber\\
  &=[\lambda_+^{\rm p1}(q_{\rm out}-q_{\rm e},q_{\rm nn})
  +\lambda_+^{\rm p2}(q_{\rm out}-q_{\rm e},q_{\rm nn})]
  P(q_{\rm out}-q_{\rm e},q_{\rm nn};t)\nonumber\\
  &+[\lambda_-^{\rm p1}(q_{\rm out}+q_{\rm e},q_{\rm nn})
  +\lambda_-^{\rm p2}(q_{\rm out}+q_{\rm e},q_{\rm nn})]P(q_{\rm out}+q_{\rm e},q_{\rm nn};t)\nonumber\\
  &+[\lambda_-^{\rm n2}(q_{\rm out},q_{\rm nn}-q_{\rm e})]
  P(q_{\rm out},q_{\rm nn}-q_{\rm e};t) \nonumber\\
  &+[\lambda_+^{\rm n2}(q_{\rm out},q_{\rm nn}+q_{\rm e})]
  P(q_{\rm out},q_{\rm nn}+q_{\rm e};t) \nonumber\\
  &+[\lambda_+^{\rm n1}(q_{\rm out}+q_{\rm e},q_{\rm nn}-q_{\rm e})]
  P(q_{\rm out}+q_{\rm e},q_{\rm nn}-q_{\rm e};t) \nonumber\\
  &+[\lambda_-^{\rm n1}(q_{\rm out}-q_{\rm e},q_{\rm nn}+q_{\rm e})]
  P(q_{\rm out}-q_{\rm e},q_{\rm nn}+q_{\rm e};t) \nonumber\\
  &-\lambda_0(q_{\rm out},q_{\rm nn})
  P(q_{\rm out},q_{\rm nn};t),
\end{align}
\vspace{-8pt}\\
where the rate $\lambda_0(q_{\rm out},q_{\rm nn})$ is the sum of all pairs of transition rates, i.e., the escape rate.
\vspace{-2pt}
\subsection{Balance of Energy and Entropy Production}
\vspace{-2pt}
Consider first the energy transfer from the isothermal environment at temperature $T$ to the system, which is interpreted as heat. Let us denote $\delta Q_\pm^\rho$ as the energy changes for a pair of transitions of a device $\rho$. By conservation of energy, for each transition, this can be written as the energy change of the entire system except for the energy change of the voltage sources. It can also be written as a sum of two terms: a conservative one and a nonconservative one. For device n1, these are
\begin{align}
  \delta Q_\pm^{\rm n1} = \Psi(q_{\rm out}\mp q_{\rm e}, q_{\rm nn}\pm q_{\rm e}) - \Psi(q_{\rm out}, q_{\rm nn}),
\end{align}
where the potential $\Psi$ is defined as
\begin{align}
  \Psi(q_{\rm out}, q_{\rm nn}) = \Phi(q_{\rm out}, q_{\rm nn}) - (q_{\rm out} + q_{\rm nn})V_{\rm ss}
\end{align}
and the function $\Phi$ is given in Eq. (\ref{eq:phi}) in the Appendix.
The other ones are given by
\vspace{-8pt}\\
\begin{align}
  \delta Q_{\pm}^{\rm n2}
  &=\Psi(q_{\rm out},q_{\rm nn}\mp q_{\rm e})-\Psi(q_{\rm out},q_{\rm nn}),\\
  \delta Q_{\pm}^{\rm p1}
  &=\Psi(q_{\rm out}\pm q_{\rm e},q_{\rm nn})-\Psi(q_{\rm out},q_{\rm nn})\mp q_{\rm e}\Delta V,\\
  \delta Q_{\pm}^{\rm p2}
  &=\Psi(q_{\rm out}\pm q_{\rm e},q_{\rm nn})-\Psi(q_{\rm out},q_{\rm nn})\mp q_{\rm e}\Delta V,
\end{align}
\vspace{-8pt}\\
where $\Delta V=V_{\rm dd}-V_{\rm ss}$. For all energy changes, the first two terms represent the conservative contribution, which means that the net change of $\Psi$ is zero for any cyclic transition starting from any state $\boldsymbol{q}$. The term $\mp q_{\rm e}\Delta V$ represents the nonconservative contribution due to the voltage difference $\Delta V$, which can be interpreted as work done by the voltage sources. The heats $\delta Q_\pm^\rho$ are also consistent with the LDB conditions, which can be confirmed by using the expression of the transition rates:
\begin{gather}
  \ln\dfrac{\lambda_+^{\rm n1}(q_{\rm out},q_{\rm nn})}{\lambda_-^{\rm n1}(q_{\rm out}-q_{\rm e},q_{\rm nn}+q_{\rm e})}
  =\beta\delta Q_+^{\rm n1},\\
  \ln\dfrac{\lambda_+^{\rm n2}(q_{\rm out},q_{\rm nn})}{\lambda_-^{\rm n2}(q_{\rm out},q_{\rm nn}-q_{\rm e})}
  =\beta\delta Q_+^{\rm n2},\\
  \ln\dfrac{\lambda_+^{\rm p1}(q_{\rm out},q_{\rm nn})}{\lambda_-^{\rm p1}(q_{\rm out}+q_{\rm e},q_{\rm nn})}
  =\beta\delta Q_+^{\rm p1},\\
  \ln\dfrac{\lambda_+^{\rm p2}(q_{\rm out},q_{\rm nn})}{\lambda_-^{\rm p2}(q_{\rm out}+q_{\rm e},q_{\rm nn})}
  =\beta\delta Q_+^{\rm p2}.
\end{gather}
where $\beta=(k_{\rm B}T)^{-1}$ is the inverse temperature. If $\Delta V =0$, for isothermal conditions, there exists an equilibrium state given by
\begin{align}
  P_{\rm eq}(q_{\rm out},q_{\rm nn}) \propto e^{-\beta\Psi(q_{\rm out},q_{\rm nn})}.
\end{align}

The heat fluxes from the environment around each device to the system are respectively given by
\begin{align}
  \langle\dot{Q}_{\rm n1}\rangle
  & = \sum_{\boldsymbol{q}}
  \delta Q_+^{\rm n1}(q_{\rm out},q_{\rm nn})(J_+^{\rm n1}(q_{\rm out},q_{\rm nn},t) \nonumber\\
  & \hspace{75pt}- J_-^{\rm n1}(q_{\rm out}-q_{\rm e},q_{\rm nn}+q_{\rm e},t)), \\
  \langle\dot{Q}_{\rm n2}\rangle
  & = \sum_{\boldsymbol{q}}
  \delta Q_+^{\rm n2}(q_{\rm out},q_{\rm nn})(J_+^{\rm n2}(q_{\rm out},q_{\rm nn},t) \nonumber\\
  & \hspace{75pt}- J_-^{\rm n2}(q_{\rm out},q_{\rm nn}-q_{\rm e},t)), \\
  \langle\dot{Q}_{\rm p1}\rangle
  & = \sum_{\boldsymbol{q}}
  \delta Q_+^{\rm p1}(q_{\rm out},q_{\rm nn})(J_+^{\rm p1}(q_{\rm out},q_{\rm nn},t) \nonumber\\
  & \hspace{75pt}- J_-^{\rm p1}(q_{\rm out}+q_{\rm e},q_{\rm nn},t)), \\
  \langle\dot{Q}_{\rm p2}\rangle
  & = \sum_{\boldsymbol{q}}
  \delta Q_+^{\rm p2}(q_{\rm out},q_{\rm nn})(J_+^{\rm p2}(q_{\rm out},q_{\rm nn},t) \nonumber\\
  & \hspace{75pt}- J_-^{\rm p2}(q_{\rm out}+q_{\rm e},q_{\rm nn},t)),
\end{align}
where $J_\pm^\rho (\boldsymbol{q},t)= \lambda_\pm^\rho(\boldsymbol{q}) P(\boldsymbol{q},t)$ are the probability flows associated with device $\rho$. For some state $\boldsymbol{q}$ and one after a possible transition from $\boldsymbol{q}$, the average electric currents through devices n1, n2, p1, and p2 are respectively given by the average stochastic forward current minus the reverse one:
\begin{align}
  \langle I _{\rm n1}\rangle_{\boldsymbol{q}} 
  &= q_{\rm e}(J_+^{\rm n1}(q_{\rm out},q_{\rm nn},t) - J_-^{\rm n1}(q_{\rm out}-q_{\rm e},q_{\rm nn}+q_{\rm e},t)),\\
  \langle I _{\rm n2}\rangle_{\boldsymbol{q}} 
  &= q_{\rm e}(J_+^{\rm n2}(q_{\rm out},q_{\rm nn},t) - J_-^{\rm n2}(q_{\rm out},q_{\rm nn}-q_{\rm e},t)),\\
  \langle I _{\rm p1}\rangle_{\boldsymbol{q}} 
  &= q_{\rm e}(J_+^{\rm p1}(q_{\rm out},q_{\rm nn},t) - J_-^{\rm p1}(q_{\rm out}+q_{\rm e},q_{\rm nn},t)),\\
  \langle I _{\rm p2}\rangle_{\boldsymbol{q}} 
  &= q_{\rm e}(J_+^{\rm p2}(q_{\rm out},q_{\rm nn},t) - J_-^{\rm p2}(q_{\rm out}+q_{\rm e},q_{\rm nn},t)).
\end{align}
Using these expressions, the heat flux associated with a device $\rho$ can be rewritten as
\begin{align}
  \langle\dot{Q}_{\rho}\rangle
  & = q_{\rm e}^{-1}\sum_{\boldsymbol{q}}\delta Q_+^{\rho}(q_{\rm out},q_{\rm nn})\langle I _{\rho}\rangle_{\boldsymbol{q}}.
\end{align}
The change in the average potential $\Psi$ is then
\begin{align}
  \dfrac{d}{dt}\langle\Psi\rangle 
  &= \langle\dot{Q}_{\rm n1}\rangle + \langle\dot{Q}_{\rm n2}\rangle + \langle\dot{Q}_{\rm p1}\rangle + \langle\dot{Q}_{\rm p2}\rangle \nonumber\\
  &\quad+\langle I_{\rm p1}\rangle\Delta V + \langle I_{\rm p2}\rangle\Delta V, \label{eq:ebalance}
\end{align}
where
\begin{align}
  \langle I_{\rm p1}\rangle=\sum_{\boldsymbol{q}}\langle I _{\rm p1}\rangle_{\boldsymbol{q}}, \quad 
  \langle I_{\rm p2}\rangle=\sum_{\boldsymbol{q}}\langle I _{\rm p2}\rangle_{\boldsymbol{q}}.
\end{align}
Equation (\ref{eq:ebalance}) thus represents the balance of energy for the system.

The entropy of the system is given by 
\begin{align}
  S=k_{\rm B}H(\boldsymbol{q})\equiv-k_{\rm B}\sum_{\boldsymbol{q}}P(\boldsymbol{q};t)\ln P(\boldsymbol{q};t).
\end{align}
The entropy flow rate, which is the change in entropy of the thermal reservoir, is given by the total heat flux from the system to the reservoir divided by the isothermal temperature, i.e.,
\begin{align}
  \dot{\Sigma}_{\rm e} = -\dfrac{1}{T}\left( \langle\dot{Q}_{\rm n1}\rangle + \langle\dot{Q}_{\rm n2}\rangle + \langle\dot{Q}_{\rm p1}\rangle + \langle\dot{Q}_{\rm p2}\rangle \right)
  \equiv -\dfrac{1}{T}\langle\dot{Q}\rangle.
\end{align}
In accordance with the master equation and the LDB conditions, the entropy production rate, which is the total entropy change of the system including the isothermal reservoir, is written as
\begin{align}
  \dot{\Sigma} 
  &\equiv \dfrac{d}{dt}S + \dot{\Sigma}_{\rm e} \\
  &=k_{\rm B}\sum_{\boldsymbol{q}}\sum_{\rho}
  \{J_-^\rho(\boldsymbol{q}+q_{\rm e}\boldsymbol{\Delta}_\rho,t) - J_+^{\rho}(\boldsymbol{q},t)\} \nonumber\\
  &\hspace{120pt}\times\ln\dfrac{J_-^\rho(\boldsymbol{q}+q_{\rm e}\boldsymbol{\Delta}_\rho,t)}{J_+^{\rho}(\boldsymbol{q},t)}, \label{eq:tep}
\end{align}
where the vector $\boldsymbol{\Delta}_\rho$ encodes the change in the state $\boldsymbol{q}$ for each transition.
By applying the log sum inequality to Eq. (\ref{eq:tep}), the entropy production rate is always nonnegative. Hence, the total entropy produced during a time interval $t\in[0,\tau]$ is also nonnegative:
\begin{align}
  \Sigma\equiv\int_0^\tau dt\,\dot{\Sigma}\geq 0.
\end{align}
This inequality represents the general second law of thermodynamics:
\begin{align}
  -\langle Q \rangle \geq -T \Delta S = -k_{\rm B}T\Delta H,
\end{align}
where
\begin{align}
  \langle Q \rangle = \int_0^\tau dt\,\langle\dot{Q}\rangle.
\end{align}
Note that we can also construct the energy balance and define the entropy functions at the trajectory level \cite{freitas2021stochastic}. By averaging these thermodynamic quantities over all possible trajectories, we can recover the usual expressions for the state ensemble.
\section{Extension of the Landauer Bound}
In order to evaluate the energetic resources of the model in view of computation, let us consider an extension of the Landauer bound from two aspects.
\vspace{2pt}
\subsection{General Landauer Bound}
\vspace{-2pt}
Let $X$ be a random variable of the microscopic states $x$ that the system can take and $M$ be the random variable of the logical states $m$. Due to the nonnegativity of the total entropy production and the chain rule of entropy \cite{wolpert2019stochastic,cover2013elements}, the heat dissipated by any computational process is bounded as follows:
\begin{align}
  -\beta \langle Q \rangle \geq -\Delta H(M) - \Delta H(X|M),
\end{align}
where $\Delta H(M)$ is the change in entropy over logical states $m$ during the time interval $t\in[0,\tau]$ and $\Delta H(X|M)$ is the change in conditional entropy over microscopic states $x$ conditioned by logical states $m$ for the same interval. The first term is often called the ``general Landauer bound" and represents the degree of logical irreversibility over logical states. The second term represents the entropy change of the internal physical degrees of freedom for the corresponding logical state. Note that the second term can be negative even if the first term is positive.
\vspace{2pt}
\subsection{Initial State Dependency and Mismatch Cost}
\vspace{-2pt}
The amount of dissipation varies according to the initial distribution over states, even for the same process. Kolchinsky and Wolpert \cite{kolchinsky2017dependence,wolpert2020thermodynamics,kolchinsky2021dependence} showed that for a fixed physical process, the total entropy production arising from the process can be decomposed into two terms:
\begin{align}
    \Sigma = -k_{\rm B}\Delta D[P(X)\|P_{\rm opt}(X)] + \Sigma^{\rm min}, \label{eq:mc}
\end{align}
where
\vspace{-9pt}\\
\begin{align}
  &\Delta D[P(X)\|P_{\rm opt}(X)] \nonumber \\
  &\hspace{30pt}=D[P(X_\tau)\|P_{\rm opt}(X_\tau)] - D[P(X_0)\|P_{\rm opt}(X_0)]
\end{align}
\vspace{-9pt}\\
is the change in the Kullback-Leibler (KL) divergence between distributions $P$ and $P_{\rm opt}$. $P(X_0)$ is any initial distribution over microscopic states $x$ and $P_{\rm opt}(X_0)$ is the initial distribution which minimizes the entropy production $\Sigma$ for a given process. $P(X_\tau)$ and $P_{\rm opt}(X_\tau)$ are the distributions at time $\tau$. $\Sigma^{\rm min}$ is the entropy produced when the initial distribution is $P_{\rm opt}(X_0)$. The first term in Eq. (\ref{eq:mc}) is called the ``mismatch cost" of running the process on the initial distribution $P$. This cost is the entropy produced due to the difference in the initial distribution from the optimal one. Due to the monotonicity of the KL divergence, the mismatch cost is always nonnegative. The minimal entropy production $\Sigma^{\rm min}$ is called the ``residual entropy production" of the process. This is also nonnegative due to the second law of thermodynamics. The optimal distribution $P_{\rm opt}$ depends on the conditional distribution for a given process and the heat $\langle Q\rangle$, which encodes various details of the physical process under consideration, such as the precise trajectory of the driving protocol \cite{kolchinsky2021dependence}. Although it is generally difficult to identify the optimal initial distribution for a given process, for the case that the process has no driving protocol, this is just the equilibrium distribution $P_{\rm eq}$ and the residual entropy production is 0.

The mismatch cost can also be defined for other thermodynamic quantities, such as nonadiabatic entropy production and free energy loss \cite{kolchinsky2021dependence}. Since there are nonconservative contributions for heat when $\Delta V \neq 0$, the entropy production can be decomposed into nonadiabatic and adiabatic components:
\begin{align}
  \Sigma = \Sigma_{\rm na} + \Sigma_{\rm a},
\end{align}
where the rates are defined as
\begin{align}
  \dot{\Sigma}_{\rm na} &= \dfrac{d}{dt}S + k_{\rm B}\sum_{\boldsymbol{q}}\dfrac{d}{dt}P(\boldsymbol{q};t)\ln P(\boldsymbol{q};t) \\
  \dot{\Sigma}_{\rm a}  &=k_{\rm B}\sum_{\boldsymbol{q}}\sum_{\rho}\left\{ J_-^{\rho}(\boldsymbol{q}+q_{\rm e}\boldsymbol{\Delta}_\rho,t) - J_+^{\rho}(\boldsymbol{q},t) \right\} \nonumber\\
  &\hspace{50pt}\times\ln
  \dfrac{\lambda_-^{\rho}(\boldsymbol{q}+q_{\rm e}\boldsymbol{\Delta}_\rho)P_{\rm ss}(\boldsymbol{q}+q_{\rm e}\boldsymbol{\Delta}_\rho;t)}{\lambda_+^{\rho}(\boldsymbol{q})P_{\rm ss}(\boldsymbol{q};t)}
\end{align}
The nonadiabatic entropy production is then written as
\begin{align}
  \Sigma_{\rm na}
    = - k_{\rm B}\Delta D[P(X)\|P_{\rm opt}(X)] + \Sigma_{\rm na}^{\rm min},
\end{align}
where the second term is the residual entropy production for the nonadiabatic one. Note that for the case that the process has a certain driving protocol, the optimal initial distribution $P_{\rm opt}$ is just the steady-state distribution $P_{\rm ss}$ and $\Sigma_{\rm na}^{\rm min}=0$.

Using the above results and the chain rule of the KL divergence, the heat dissipated by  any process can be written as
\vspace{-9.5pt}\\
\begin{align}
  -\beta\langle Q \rangle 
  &= - \Delta H(M) - \Delta D[P(M)\|P_{\rm opt}(M)] \nonumber\\
  &\quad - \Delta H(X|M) - \Delta D[P(X|M)\|P_{\rm opt}(X|M)]\nonumber \\
  &\quad+ \Sigma_{\rm na}^{\rm min}/k_{\rm B} + \Sigma_{\rm  a}/k_{\rm B}.
\end{align}
\vspace{-9.5pt}\\
where the second term is the difference in the KL divergence over logical states and the fourth term is the difference in the conditional KL divergence, which is also nonnegative. Here, we refer to the second term as the ``logical mismatch cost". In the following, we will focus on the general Landauer bound and the logical mismatch cost.

\section{Results}
\subsection{Stationary State}
\begin{figure*}[ht]
  \vspace{12pt}
  \centering
  \begin{minipage}[b]{0.45\textwidth}
    \centering
    \leftline{\footnotesize (a)}
    \vspace{3pt}
    \includegraphics[scale=0.55]{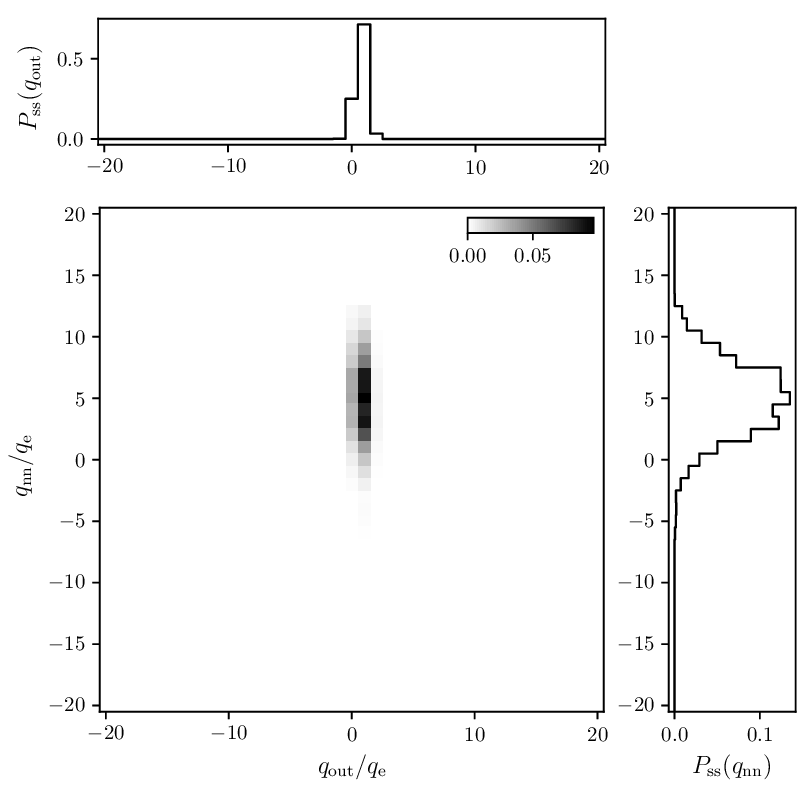}
  \end{minipage}
  \begin{minipage}[b]{0.45\textwidth}
    \centering
    \leftline{\footnotesize (b)}
    \vspace{3pt}
    \includegraphics[scale=0.55]{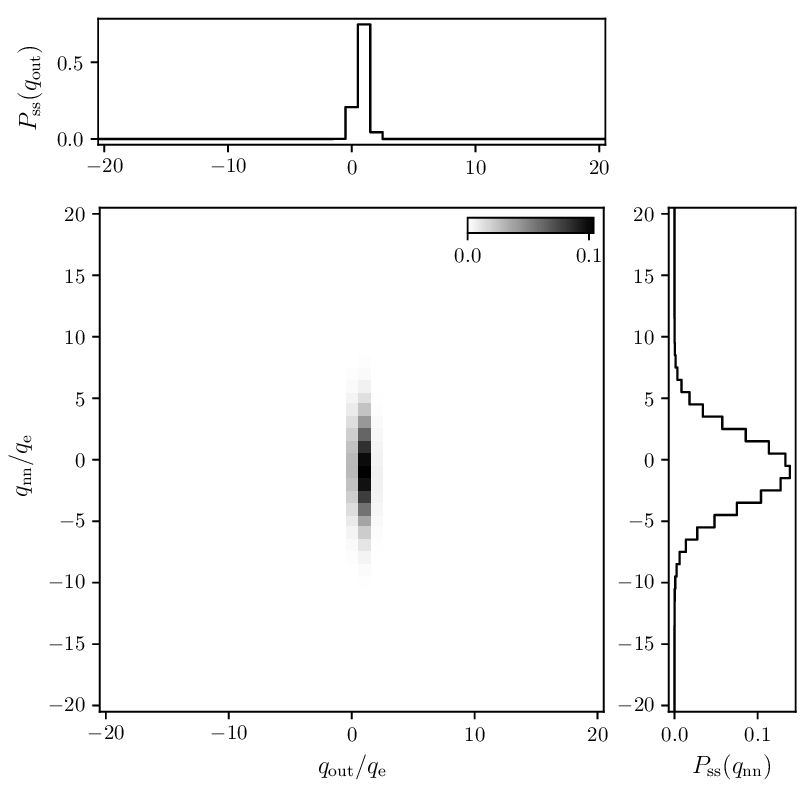}
  \end{minipage}
  \begin{minipage}[b]{0.45\textwidth}
    \centering
    \leftline{\footnotesize (c)}
    \vspace{3pt}
    \includegraphics[scale=0.55]{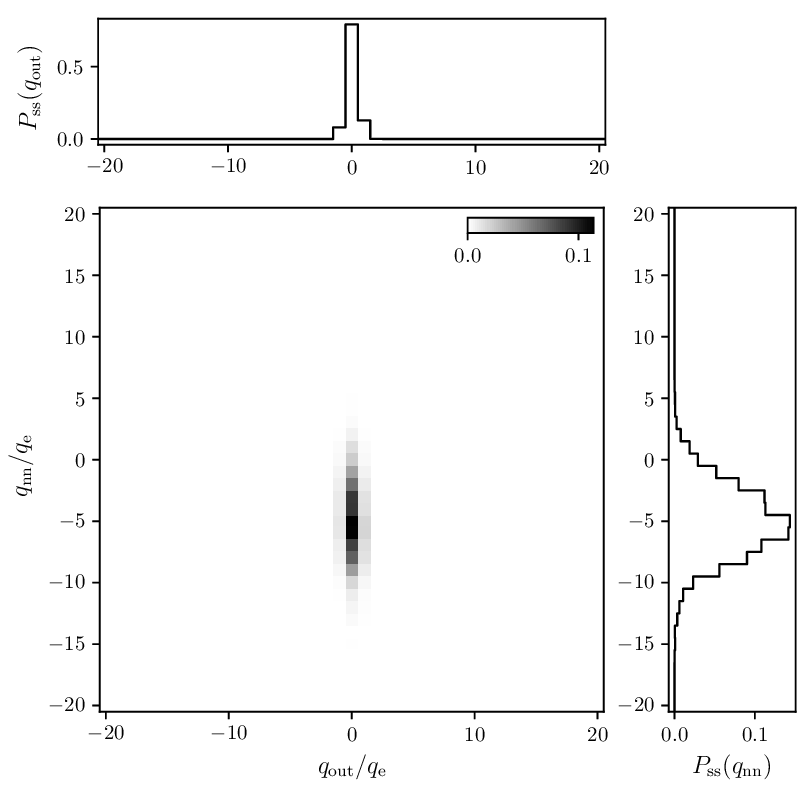}
  \end{minipage}
  \begin{minipage}[b]{0.45\textwidth}
    \centering
    \leftline{\footnotesize (d)}
    \vspace{3pt}
    \includegraphics[scale=0.55]{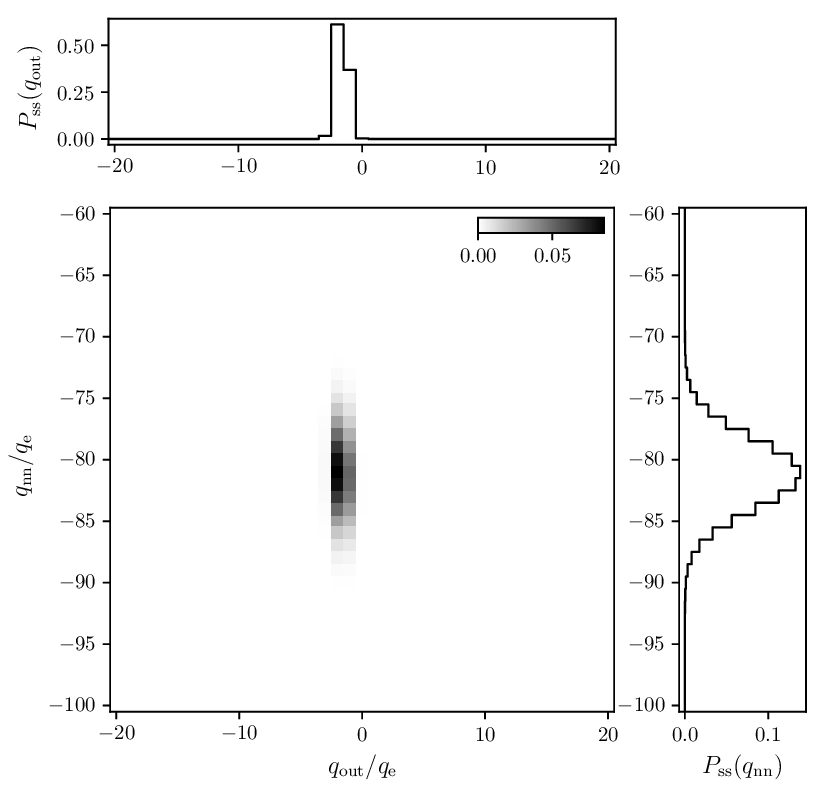}
  \end{minipage}
  \caption{Estimated steady-state distribution $P_{\rm ss}(q_{\rm out},q_{\rm nn})$ for four pairs of input voltages, (a) $(V_{\rm in,A}/V_{\rm T},V_{\rm in,B}/V_{\rm T})=(-5,-5)$, (b) $(V_{\rm in,A}/V_{\rm T},V_{\rm in,B}/V_{\rm T})=(-5,+5)$, (c) $(V_{\rm in,A}/V_{\rm T},V_{\rm in,B}/V_{\rm T})=(+5,-5)$, and (d) $(V_{\rm in,A}/V_{\rm T},V_{\rm in,B}/V_{\rm T})=(+5,+5)$. The two-dimensional color maps present the steady-state distribution $P_{\rm ss}(q_{\rm out},q_{\rm nn})$ and the upper left and bottom right figures for each pair of figures respectively show the marginal distribution $P_{\rm ss}(q_{\rm out})$ and $P_{\rm ss}(q_{\rm nn})$. For all pairs, $V_{\rm dd}=-V_{\rm ss}=5V_{\rm T}$, $V_{\rm T}=26\,$mV, $C_{\rm g}=50\,$aF, $r=100$ and $n=1$.}\label{fig:SSD}
\end{figure*}

\vspace{-2pt}
Since the dynamics of the system are ergodic, there exists a unique steady-state distribution $P_{\rm ss}$. If the state space is finite, the distribution can be obtained by calculating the eigenvector of the time-homogeneous transition rate matrix of the master equation with zero eigenvalue. However, if the state space is infinitely countable, such a method does not work in general. One way to estimate the steady-state distribution for this case is to sample a single trajectory of the state over a long time interval and then calculate the time window of the remaining trajectories in each of the observed states \cite{kuntz2021stationary}. For a given long time interval $\tau_0$, the steady-state probability for state $\boldsymbol{q}$ is approximately expressed by
\vspace{-7pt}\\
\begin{align}
  P_{\rm ss}(\boldsymbol{q})\approx\dfrac{t(\boldsymbol{q})}{\tau_0},
\end{align}
\vspace{-7pt}\\
where $t(\boldsymbol{q})\in[0,\tau_0)$ is the total time spent in the state $\boldsymbol{q}$. We can use the Gillespie stochastic simulation algorithm (GSSA) to sample a single trajectory for four pairs of input voltages, namely, $(V_{\rm in,A}/V_{\rm T},V_{\rm in,B}/V_{\rm T})=(-5,-5)$, $(-5,+5)$, $(+5,-5)$ and $(+5,+5)$. The associated parameters are chosen as $V_{\rm dd}/V_{\rm T}=-V_{\rm ss}/V_{\rm T}=5$, $V_{\rm T}=26\,$mV (room temperature), $C_{\rm g}=50\,$aF, $r=100$ and $n=1$. Figure \ref{fig:SSD} shows the estimated steady-state distributions $P_{\rm ss}(q_{\rm out},q_{\rm nn})$ for four pairs of input voltages. The total time interval $\tau_0$ is such that it is much longer than the global time scale $t_0$ in all cases.

In order for the system to function as a NAND gate, the resulting logical states over each steady-state distribution should reflect the truth table of the gate. Here, we define the logical states as sets of microscopic states $\boldsymbol{q}$ conditioned by the output voltages $V_{\rm out}$, i.e.,
\begin{align}
  {\rm H}&=\left\{\boldsymbol{q}\,\middle|\,V_{\rm out}(\boldsymbol{q})>\dfrac{V_{\rm dd}+V_{\rm ss}}{2}\right\},\label{eq:high}\\
  {\rm L}&=\left\{\boldsymbol{q}\,\middle|\,V_{\rm out}(\boldsymbol{q})\leq\dfrac{V_{\rm dd}+V_{\rm ss}}{2}\right\},\label{eq:low}
\end{align}
where H and L correspond to the HIGH and LOW logical states. Hence, the coarse-grained steady-state distribution can be written as $P_{\rm ss}(V_{\rm out}>(V_{\rm dd}+V_{\rm ss})/2)$ for the HIGH state and $P_{\rm ss}(V_{\rm out}\leq (V_{\rm dd}+V_{\rm ss})/2)$ for the LOW state.
\begin{table}[h]
  \caption{Coarse-grained distributions over the estimated steady-state distribution $P_{\rm ss}(q_{\rm out},q_{\rm nn})$. The values in brackets are statistical uncertainties.}
  \label{tab:truth}
  \vspace{3pt}
  \footnotesize
  \begin{tabularx}{\columnwidth}{@{} >{\centering\arraybackslash}X>{\centering\arraybackslash}X>{\centering\arraybackslash}X>{\centering\arraybackslash}X @{}}
      \hline
      $V_{\rm in,A}/V_{\rm T}$ & $V_{\rm in,B}/V_{\rm T}$ & $P_{\rm ss}(V_{\rm out}>0)$ & $P_{\rm ss}(V_{\rm out}\leq 0)$ \\
      \hline
      -5 & -5 &  0.9986(1) &  0.0014(1) \\
      -5 & +5 & 0.99907(4) & 0.00093(4) \\
      +5 & -5 & 0.99986(1) & 0.00014(1) \\
      +5 & +5 & 0.00372(1) & 0.99628(1) \\
      \hline
  \end{tabularx}
\end{table}
Table \ref{tab:truth} shows the resulting coarse-grained distributions over the estimated steady-state distribution $P_{\rm ss}(\boldsymbol{q})$ for each pair of input voltages. For all pairs, the result shows that the output voltage $V_{\rm out}$ in the steady state brings the expected computational result stochastically within an error of less than 0.4\% if we define the logical states as Eqs. (\ref{eq:high}) and (\ref{eq:low}).

\subsection{Analysis of the Lower Bound}\label{subsec:4-2}
Using the NAND gate model, we can analyze the generalized Landauer bound and the logical
mismatch cost. We first have to know the optimal initial distribution $P_{\rm opt}$ to evaluate the logical mismatch cost, and it depends on the physical details of the system and the driving protocol. Here, we consider a computational process as follows:
\begin{enumerate}
  \item At time $t < 0$, $V_{\rm dd} = V_{\rm ss} = 0$, which means no power is supplied to the system. The input voltages $V_{\rm in,A}$ and $V_{\rm in,B}$ are set to certain values.
  \item Immediately before $t = 0$, the voltages $V_{\rm dd}$ and $V_{\rm ss}$ are instantaneously changed as $V_{\rm dd}=-V_{\rm ss}=5V_{\rm T}$. This is realized by the quench process.
  \item At time $t > 0$, all external voltages are fixed. This is a relaxation process to the (nonequilibrium) steady state.
  \item At time $t = t_0$, which is expected to be the global relaxation time scale, one recognizes the observed state as the computational result of the NAND gate via the output voltage $V_{\rm out}$.
\end{enumerate}
When $V_{\rm dd} = V_{\rm ss}$, the system is in the equilibrium state given by $P_{\rm eq}(\boldsymbol{q}) \propto e^{-\beta \Psi(\boldsymbol{q})}$. In particular, when $V_{\rm dd} = V_{\rm ss}=0$, the equilibrium distribution can be written as
\begin{align}
  P_{\rm eq}(\boldsymbol{q})
  \propto\exp{\left(-\dfrac{1}{2}(\boldsymbol{q}-\boldsymbol{\mu})^{\rm T}
  \boldsymbol{\Sigma}^{-1}
  (\boldsymbol{q}-\boldsymbol{\mu})
  \right)},
\end{align}
where
\begin{align}
  \boldsymbol{\mu}
  =\begin{pmatrix}
    0 \\
    -C_{\rm g}V_{\rm in,A}
  \end{pmatrix}\quad {\rm and}\quad 
  \boldsymbol{\Sigma}=\dfrac{q_{\rm e}q_{\rm T}}{3r+5}
  \begin{pmatrix}
    3 & -1 \\
    -1 & 2+r
  \end{pmatrix}.
\end{align}
This distribution is a discrete analogue to the bivariate normal distribution with mean $\boldsymbol{\mu}$ and covariance matrix $\boldsymbol{\Sigma}$. Since the system relaxes to a steady state after the quench process, the optimal initial distribution is just the steady-state distribution $P_{\rm ss}$, which results in zero residual nonadiabatic entropy production. Thus, the dissipated heat can be expressed as
\begin{align}
  -\beta\langle Q \rangle 
  &= -\Delta H(M) - \Delta D[P(M)\|P_{\rm ss}(M)] \nonumber\\
  &\quad -\Delta H(X|M) - \Delta D[P(X|M)\|P_{\rm ss}(X|M)]\nonumber \\
  &\quad + \Sigma_{\rm  a}/k_{\rm B}. \label{eq:finalex}
\end{align}
To analyze the general lower bound and the logical mismatch cost, we have to sample the initial distribution $P_{\rm eq}$ and the final one. Since the initial distribution is analogous to a Gaussian one, we obtain the initial states by sampling real values from a bivariate normal distribution and then rounding around $\pm q_{\rm e}/2$. Using the GSSA, the final distribution can be obtained by sampling the trajectories starting from these initial states. Since each trajectory can be sampled independently, we decided to create a program written in CUDA C/C++ to exploit the parallel processing capability of GPUs. Table \ref{tab:inidist} and \ref{tab:findist} show the coarse-grained initial and final distributions based on the definitions of the logical states for different pairs of input voltages. In all cases, we sampled $10^5$ independent trajectories by using the GSSA. Note that there is no truncation of the state space.
\begin{table}[h]
  \caption{Coarse-grained initial distributions for different pairs of input voltages.}
  \label{tab:inidist}
  \vspace{3pt}
  \footnotesize
  \begin{tabularx}{\columnwidth}{>{\centering\hsize=.15\hsize\linewidth=\hsize}X>{\centering\hsize=.15\hsize\linewidth=\hsize}X>{\centering\hsize=.25\hsize\linewidth=\hsize}X>{\centering\arraybackslash\hsize=.25\hsize\linewidth=\hsize}X}
      \hline
      $V_{\rm in,A}/V_{\rm T}$ & $V_{\rm in,B}/V_{\rm T}$ & $P(V_{\rm out}>0)$ & $P(V_{\rm out}\leq 0)$ \\
      \hline
      $-5$ & $-5$ & 0.8466 & 0.1534 \\
      $-5$ & $+5$ & 0.8462 & 0.1538 \\
      $+5$ & $-5$ & 0.8486 & 0.1514 \\
      $+5$ & $+5$ & 0.8446 & 0.1554 \\
      \hline
  \end{tabularx}
  \vspace{-48pt}
\end{table}
\begin{table}[h]
  \caption{Coarse-grained final distributions for different pairs of input voltages.}
  \label{tab:findist}
  \vspace{3pt}
  \footnotesize
  \begin{tabularx}{\columnwidth}{>{\centering\hsize=.15\hsize\linewidth=\hsize}X>{\centering\hsize=.15\hsize\linewidth=\hsize}X>{\centering\hsize=.25\hsize\linewidth=\hsize}X>{\centering\arraybackslash\hsize=.25\hsize\linewidth=\hsize}X}
      \hline
      $V_{\rm in,A}/V_{\rm T}$ & $V_{\rm in,B}/V_{\rm T}$ & $P(V_{\rm out}>0)$ & $P(V_{\rm out}\leq 0)$ \\
      \hline
      $-5$ & $-5$ & 0.9951 & 0.0049 \\
      $-5$ & $+5$ & 0.9999 & 0.0001 \\
      $+5$ & $-5$ & 0.9804 & 0.0196 \\
      $+5$ & $+5$ & 0.0112 & 0.9888 \\
      \hline
  \end{tabularx}
\end{table}

The general Landauer bound for the process is obtained from the results in Table \ref{tab:inidist} and \ref{tab:findist}. By applying the estimated steady-state distributions, the logical mismatch cost is also obtained for different pairs of input voltages. Table \ref{tab:GLBandLMC} shows the numerical results for these two costs for that process.
\begin{table}[hb]
  \vspace{-12pt}
  \caption{General Landauer bound and logical mismatch cost for different pairs of input voltages.}
  \label{tab:GLBandLMC}
  \vspace{3pt}
  \footnotesize
  \begin{tabularx}{\columnwidth}{>{\centering\hsize=.15\hsize\linewidth=\hsize}X>{\centering\hsize=.15\hsize\linewidth=\hsize}X>{\centering\hsize=.3\hsize\linewidth=\hsize}X>{\centering\arraybackslash\hsize=.4\hsize\linewidth=\hsize}X}
      \hline
      $V_{\rm in,A}/V_{\rm T}$ & $V_{\rm in,B}/V_{\rm T}$ & $-k_{\rm B}T\Delta H(M)$ & $-k_{\rm B}T\Delta D[P(M)\|P_{\rm ss}(M)]$ \\
      \hline
      -5 & -5 & $1.656\times 10^{-21}\,$J & $2.408\times 10^{-21}\,$J \\
      -5 & +5 & $1.784\times 10^{-21}\,$J & $2.684\times 10^{-21}\,$J \\
      +5 & -5 & $1.369\times 10^{-21}\,$J & $3.503\times 10^{-21}\,$J \\
      +5 & +5 & $1.535\times 10^{-21}\,$J & $1.787\times 10^{-20}\,$J \\
      \hline
  \end{tabularx}
\end{table}
For all pairs of inputs, the general Landauer bound has order $10^{-21}\,$J, which is the same order as the original Landauer bound at room temperature. The logical mismatch cost also has the order of the general Landauer bounds, except for the case $(V_{\rm in,A}, V_{\rm in,B}) = (5V_{\rm T}, 5V_{\rm T})$. For this case, it is ten times larger than the others, because the distribution of output voltages is initially skewed positive, i.e., $V_{\rm out} > 0$, which results in a larger KL divergence at $t=0$.

\subsection{Relation between Dissipation, Time Scale, and Reliability of Computations}
\begin{figure}[ht]
  \vspace{12pt}
  \centering
  \begin{minipage}[b]{0.9\columnwidth}
    \centering
    \vspace{3pt}
    \includegraphics[scale=0.72]{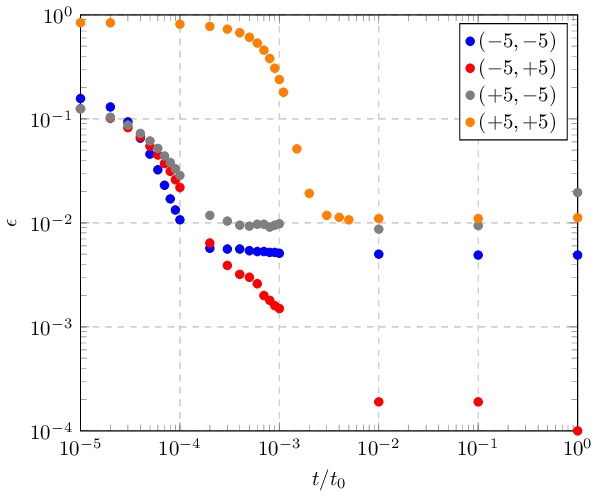}
  \end{minipage}
  \caption{Time dependence of the probability error $\epsilon$ for different pairs of input voltages $(V_{\rm in,A},V_{\rm in,B})=(-5,-5)$, $(-5,+5)$, $(+5,-5)$, and $(+5,+5)$ until time $t=t_0$.}\label{fig:ERR_T}
\end{figure}

We also investigated the relation between the dissipation, time scale, and reliability of the NAND process. Fig. \ref{fig:ERR_T} presents the time evolution of the probability error $\epsilon$, which is defined as the probability of obtaining an undesired output voltage for each pair of input voltages $(V_{\rm in,A}, V_{\rm in,B})$.
For the cases $(V_{\rm in,A}, V_{\rm in,B})=(-5,-5)$, $(-5,+5)$, and $(+5,-5)$, the behaviors of $\epsilon$ are similar until $10^{-4}$ (in units of $t_0$). After this, the three trajectories branch off, and the reliability further improves especially for $(-5,+5)$. For $(+5, +5)$, convergence of $\epsilon$ is delayed compared with the other cases. This is because the initial distribution is far from the steady-state distribution. 

\begin{figure*}[ht]
  \vspace{12pt}
  \centering
  \begin{minipage}[b]{0.45\textwidth}
    \centering
    \leftline{\footnotesize (a)}
    \vspace{3pt}
    \includegraphics[scale=0.72]{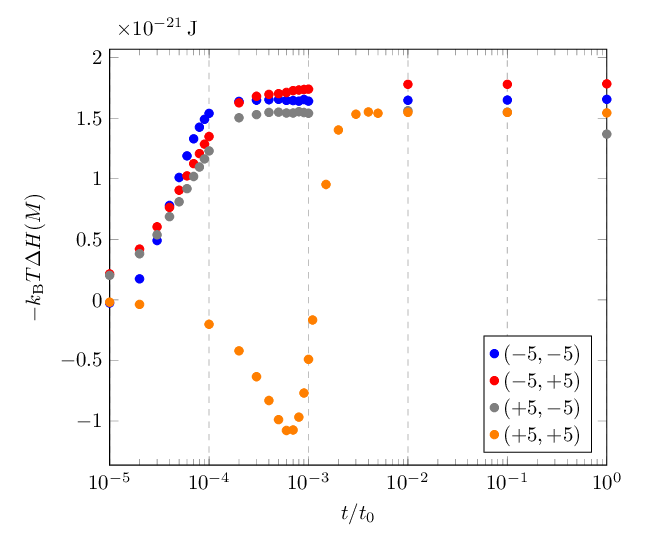}
  \end{minipage}
  \begin{minipage}[b]{0.45\textwidth}
    \centering
    \leftline{\footnotesize (b)}
    \vspace{3pt}
    \includegraphics[scale=0.72]{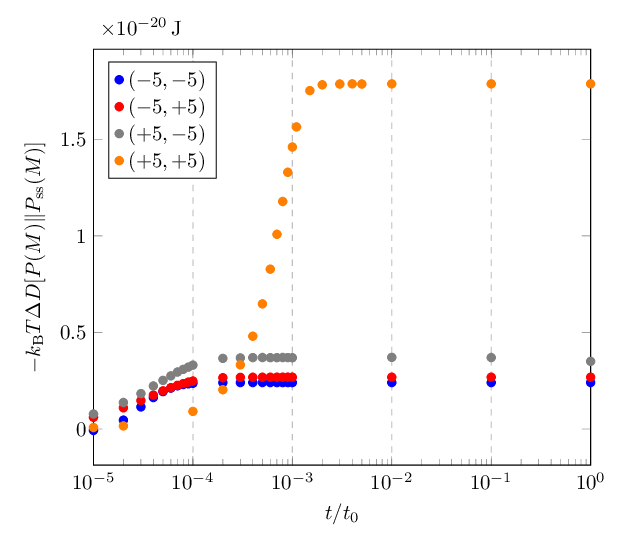}
  \end{minipage}
  \caption{Time dependence of (a) the general Landauer bound and (b) the logical mismatch cost for different pairs of input voltages $(V_{\rm in,A},V_{\rm in,B})=(-5,-5)$, $(-5,+5)$, $(+5,-5)$ and $(+5,+5)$ until time $t=t_0$.}\label{fig:HEAT_T}
\end{figure*}

Figure \ref{fig:HEAT_T} presents the time dependence of the general Landauer bound and the logical mismatch cost for different inputs.
The general Landauer bound for the $(+5,+5)$ case is very different from the others in that the bound takes a negative value during $t/t_0 = 10^{-4}\sim 10^{-3}$. However, it eventually becomes positive during $t/t_0 = 10^{-3}\sim 10^{-2}$. Thus, for that process and for all pairs of inputs, we can say that the computations show the logical irreversibility. The time dependence of the logical mismatch cost for the cases $(V_{\rm in,A},V_{\rm in,B})=(-5,-5)$, $(-5,+5)$, and $(+5,-5)$ shows the same behavior as that of the general Landauer bound. For $(+5, +5)$, however, this cost is drastically increased because the initial distribution of the output voltage is far from the steady state of that case. As a result, the sum of the general Landauer bound and the logical mismatch cost is positive most of the time.

\begin{figure*}[ht]
  \centering
  \begin{minipage}[b]{0.45\textwidth}
    \centering
    \leftline{\footnotesize (a)}
    \vspace{3pt}
    \includegraphics[scale=0.72]{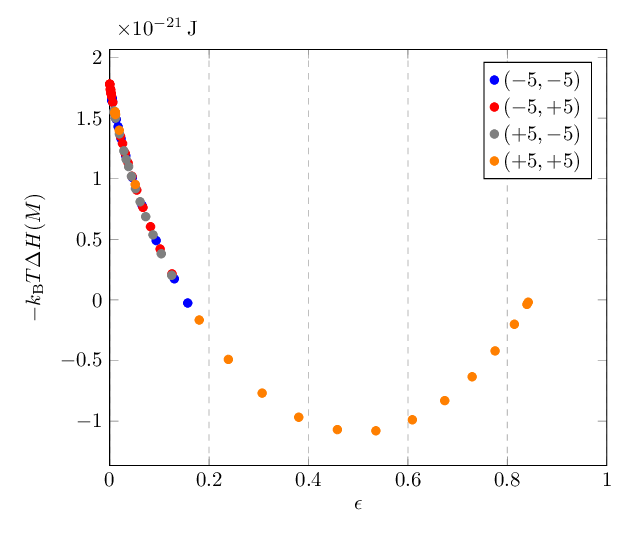}
  \end{minipage}
  \begin{minipage}[b]{0.45\textwidth}
    \centering
    \leftline{\footnotesize (b)}
    \vspace{3pt}
    \includegraphics[scale=0.72]{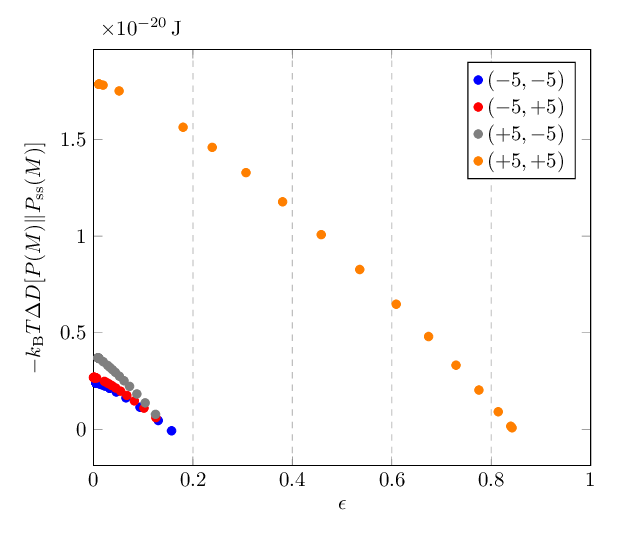}
  \end{minipage}
  \caption{Relation between error and (a) the general Landauer bound and (b) the logical mismatch cost for different pairs of input voltages $(V_{\rm in,A},V_{\rm in,B})=(-5,-5)$, $(-5,+5)$, $(+5,-5)$ and $(+5,+5)$.}\label{fig:HEAT_ERR}
  \vspace{-12pt}
\end{figure*}

Fig. \ref{fig:HEAT_ERR}(a) shows the relation between the error $\epsilon$ and the general Landauer bound.
Since the initial course-grained distribution of the output voltage $V_{\rm out}$ is almost the same for all pairs of input voltages, the relation between the general Landauer bound and the error can be explained in terms of a function of $\epsilon$, defined as
\vspace{-9pt}\\
\begin{align}
  &k_{\rm B}T (-P_{\rm eq}(0)\ln P_{\rm eq}(0) - P_{\rm eq}(1) \ln P_{\rm eq}(1) \nonumber\\
  &\hspace{75pt}+ \epsilon\ln\epsilon + (1 - \epsilon) \ln(1 - \epsilon)),
\end{align}
\vspace{-9pt}\\
where the initial course-grained probability $P_{\rm eq}(0) = 1-P_{\rm eq}(1)$ is about 0.15. If the initial distribution is uniform and the process is completely reliable, i.e., $\epsilon = 0$, then the above function reduces to the original Landauer bound $k_{\rm B}T\ln 2$. Fig. \ref{fig:HEAT_ERR}(b) presents the relation between the logical mismatch cost and the error. For all cases, as $\epsilon \rightarrow 0$, the logical mismatch cost converges to certain values. Although this result appears to be different from the well-known error-dissipation scaling $k_{\rm B} \ln(1/\epsilon)$ discussed in Refs. \citen{brillouin2013science} and \citen{riechers2020balancing}, this is not the case because the KL divergence is bounded due to the full support of the steady-state distribution. For $(+5, +5)$, this ``error-mismatch cost" relation deviates from the other cases. This is because the optimal initial distribution is different from the others.

\section{Discussion}
Logical reversibility means that we can precisely estimate the input from the output \cite{sagawa2014thermodynamic}. However, the numerical results of the general Landauer bound are just the difference in entropy over the output logical states, i.e., $V_{\rm out}$, and they don't have such implication. Hence, we need to take the input logical states into account. Since the inputs are given in the form of driving parameters, i.e., input voltages $V_{\rm in,A}$ and $V_{\rm in,B}$ and these are certain during the interval $[0,\tau]$, both the inputs and the outputs exist. Therefore, the computational process is logically reversible in spite of the positive general Landauer bound. If we want to measure the logical irreversibility via the bound, we should model a memory system about inputs and then carry out the erasure process on the memory.

In order to evaluate the energy cost in view of computation, it is more important to discuss what properties of computation are reflected in the logical mismatch cost. The original mismatch cost is based on the distance between actual initial distribution and the optimal initial distribution for a given process and nothing is implicated about the properties of computation such as logical reversibility. However, in our formulation, since the steady-state distribution is the optimal one and its coarse-grained distribution mostly reflects the correct computational result of the NAND gate, the KL divergence $D[P(M)\|P_{\rm ss}(M)]$ quantifies the distance from the correct output logical states. Considering that the initial output distribution is skewed toward the HIGH state and that the resulting logical mismatch cost for (HIGH,HIGH) inputs is ten times larger than the others, this cost could be interpreted as the $flipping$ cost of the output logical state. For (HIGH,HIGH) inputs, the output logical state mostly switches from the HIGH to LOW state, which clearly leads to a large KL divergence. This implication has a totally different meaning than that of the Landauer bound. Note as well that, in Ref. \citen{wolpert2020thermodynamics}, the mismatch cost plays an important role in analyzing the energy dissipated in the circuit systems.

Both the general Landauer bound and the logical mismatch cost are characterized by only the output voltage $V_{\rm out}$, which is experimentally accessible. If the supply voltage $\Delta V$ is small and the time interval is short to the extent that the other costs in Eq. (\ref{eq:finalex}) are negligible, then the dissipated heat $\langle Q \rangle$ is characterized by only the output voltage. However, contributions of the internal degrees of freedom for the logical states or the adiabatic entropy production cannot generally be avoided. These contributions should be characterized by the intrinsic physical quantities of the system. Characterizing these conditional costs and the adiabatic entropy production is a future work.

One interesting point about the relation between reliability and energy cost is that for different inputs, a more reliable computation doesn't always lead to a large dissipated heat. As an example, for (LOW,HIGH) inputs, the reliability of the output is much better than that of the other inputs, but the sum of the Landauer bound and the mismatch cost is not the largest. The cause of this situation is connected with the definition of the logical states, the steady-state distribution for a given input voltage, and the driving parameters. Appropriately evaluating how a difference in reliability affects the energy cost is also a future work.

\section{Conclusion}
We modeled a CMOS NAND gate operating in the sub-threshold region and analyzed the heat dissipated by its relaxation to a nonequilibrium steady state starting from instantaneous driving of power-supply voltage for different pairs of input voltages. We analyzed the heat from two aspects: the general Landauer bound and the logical mismatch cost. For this relaxation process, we found that the general Landauer bound, which represents the degree of the logical irreversibility, is the same order for different inputs. For the logical mismatch cost, we found that its order for the (HIGH, HIGH) input case is about ten times larger than the other cases due to the difference between the initial distribution and the steady-state one.
We claimed that this cost can be interpreted as the cost of flipping the logical state, which is different from the Landauer bound. We also analyzed the interplay between the costs and time scale and reliability of this process. We found that for different inputs, a more reliable computation doesn't always induce a larger cost. Our work provides a way of analyzing the thermodynamic costs of physical computational systems.\\
\begin{acknowledgments}
This work was supported in part by a JSPS Grant-in-Aid for Scientific Research (B) (JP20H01827).
\end{acknowledgments}
\appendix
\section{Model Setup}
In the model shown in Fig. \ref{fig:cmosnand}(c), there are two $free$ conductors whose charges thermodynamically fluctuate. One is associated with the output voltage $V_{\rm out}$ and the other is located between two-terminal devices n1 and n2. Besides these, there are four conductors with fixed voltages, $V_{\rm in,A}, V_{\rm in,B}, V_{\rm dd},$ and $V_{\rm ss}$. Hence, the total number of conductors is six. The charge vector and the voltage vector are respectively given by
\begin{align}
  \boldsymbol{q}_0
  = (q_{\rm out}, q_{\rm nn}, q_{\rm in,A}, q_{\rm in,B}, q_{\rm dd}, q_{\rm ss})^{\rm T}
\end{align}
and
\begin{align}
  \boldsymbol{V}_0 
  = (V_{\rm out}, V_{\rm nn}, V_{\rm in,A}, V_{\rm in,B}, V_{\rm dd}, V_{\rm ss})^{\rm T}.
\end{align}
These vectors are linearly related as
\begin{align}
  \boldsymbol{q}_0=\boldsymbol{C}_0\boldsymbol{V}_0,
\end{align}
where the capacitance matrix $\boldsymbol{C}_0$ is
\vspace{-8.5pt}\\
\begin{align}
  \begin{pmatrix}
    3C_{\rm o} & -C_{\rm o} & 0 & 0 & -2C_{\rm o} & 0 \\
    -C_{\rm o} & 2C_{\rm o}+C_{\rm g} & -C_{\rm g} & 0 & 0 & -C_{\rm o} \\
    0 & -C_{\rm g} & 2C_{\rm g} & 0 & -C_{\rm g} & 0 \\
    0 & 0 & 0 & 2C_{\rm g} & -C_{\rm g} & -C_{\rm g} \\
    -2C_{\rm o} & 0 & -C_{\rm g} & -C_{\rm g} & 2(C_{\rm o}+C_{\rm g}) & 0 \\
    0 & -C_{\rm o}& 0 & -C_{\rm g} & 0 & C_{\rm o}+C_{\rm g}
  \end{pmatrix}.
\end{align}
\vspace{-8.5pt}\\
The electrostatic energy of the system plus the voltage sources is then
\vspace{-7pt}\\
\begin{align}
  E(\boldsymbol{q})
  =\dfrac{1}{2C}
  \left[(2+r)q_{\rm out}^2
  +2q_{\rm out}q_{\rm nn}
  +3q_{\rm nn}^2\right]
  +{\rm const.},
\end{align}
\vspace{-7pt}\\
where the last term does not depend on the free charge vector $\boldsymbol{q}$. Hence, the potential $\Phi$, which is defined in Eq. (28) in Ref. \citen{freitas2021stochastic}, can be calculated as \\
\begin{strip}
  \vbox{
    \begin{align}
      \Phi(\boldsymbol{q})
      &=E(\boldsymbol{q}) 
      -
      (
        V_{\rm in,A}, 
        V_{\rm in,B}, 
        V_{\rm dd}, 
        V_{\rm ss}
      )
      \begin{pmatrix}
        0 & -C_{\rm g} \\
        0 & 0 \\
        -2C_{\rm o} & 0 \\
        0 & -C_{\rm o}
      \end{pmatrix}
      \dfrac{1}{C_{\rm o}(5C_{\rm o}+3C_{\rm g})}
      \begin{pmatrix}
        2C_{\rm o}+C_{\rm g} & C_{\rm o} \\
        C_{\rm o} & 3C_{\rm o}
      \end{pmatrix}
      \begin{pmatrix}
        q_{\rm out} \\
        q_{\rm nn}
      \end{pmatrix}\nonumber\\
      &=\dfrac{1}{2C}((2+r)q_{\rm out}^2 + 2q_{\rm out}q_{\rm nn} + 3q_{\rm nn}^2) 
      + \dfrac{2C_{\rm o}V_{\rm dd}}{C}((2+r)q_{\rm out}+q_{\rm nn})
      +\dfrac{C_{\rm o}(rV_{\rm in,A}+V_{\rm ss})}{C}(q_{\rm out}+3q_{\rm nn})
      + {\rm const}.\label{eq:phi}
    \end{align}
  }
\end{strip}

\bibliography{reference}
\bibliographystyle{jpsj}

\end{document}